\documentclass[aps,twocolumn,preprintnumbers,nobalancelastpage,superscriptaddress,nofootinbib]{revtex4}

\usepackage[english]{babel}
\usepackage{amsmath,amssymb}
\usepackage[dvips]{graphicx}
\usepackage[sort&compress]{natbib}
\usepackage{bbm}
\usepackage{color}
\usepackage{footnote}
\usepackage[normalem]{ulem}

\begin{document}

\preprint{YITP-SB-13-009}

\title{Determining Triple Gauge Boson Couplings from Higgs Data}
\author{Tyler Corbett}
\email{corbett.ts@gmail.com}
\affiliation{%
  C.N.~Yang Institute for Theoretical Physics, SUNY at Stony Brook,
  Stony Brook, NY 11794-3840, USA}

\author{O.\ J.\ P.\ \'Eboli}
\email{eboli@fma.if.usp.br}
\affiliation{Instituto de F\'{\i}sica,
             Universidade de S\~ao Paulo, S\~ao Paulo -- SP, Brazil.}

\author{J.\ Gonzalez--Fraile}
\email{fraile@ecm.ub.edu}
\affiliation{%
  Departament d'Estructura i Constituents de la Mat\`eria and
  ICC-UB, Universitat de Barcelona, 647 Diagonal, E-08028 Barcelona,
  Spain}

\author{M.\ C.\ Gonzalez--Garcia} 
\email{concha@insti.physics.sunysb.edu}
\affiliation{%
  C.N.~Yang Institute for Theoretical Physics, SUNY at Stony Brook,
  Stony Brook, NY 11794-3840, USA}
\affiliation{%
  Departament d'Estructura i Constituents de la Mat\`eria and
  ICC-UB, Universitat de Barcelona, 647 Diagonal, E-08028 Barcelona,
  Spain}
\affiliation{%
  Instituci\'o Catalana de Recerca i Estudis Avan\c{c}ats (ICREA),}

\begin{abstract}

In the framework of effective Lagrangians with the $SU(2)_L \times
U(1)_Y$ symmetry linearly realized, modifications of the couplings of the Higgs
field to the electroweak gauge bosons are related to anomalous triple
gauge couplings (TGCs). Here, we show that the analysis of the
latest Higgs boson production data at the LHC and Tevatron
give rise to strong bounds on TGCs that are complementary to those from
direct TGC analysis.  We present the constraints on TGCs obtained 
by combining  all available data on direct TGC studies and 
on Higgs production analysis.

\vskip .5cm {\bf Note added:} The analysis has been updated with all
  the public data available as November 2013. Updates of this analysis
  are provided at the web site \url{http://hep.if.usp.br/Higgs} as
  well as new versions of this manuscript.
\end{abstract}

\maketitle

\renewcommand{\baselinestretch}{1.15}


The direct exploration of the electroweak symmetry breaking sector has
recently started with the discovery of a state that resembles the
standard model (SM) Higgs boson~\cite{ Englert:1964et} at the CERN
Large Hadron Collider (LHC)~\cite{discovery}. With the increase of
available data on this Higgs-like state we can scrutinize its
couplings to determine if it is indeed the state predicted by the
SM~\cite{us1,usnew,couplings, Giardino:2013bma}.  The observation of
departures from the SM predictions for the Higgs couplings can give
hints of physics beyond the SM characterized by an energy scale
$\Lambda$.

A model independent way to parametrize the low--energy effects of
possible SM extensions is by the means of an effective
Lagrangian~\cite{effective}, which depends on the low--energy particle
content and symmetries.  This bottom--up approach has the advantage of
minimizing the amount of theoretical hypothesis when studying
deviations from the SM predictions~\cite{usnew}.  The absence of direct 
new physics (NP) signals in the present LHC runs so far and the observation of 
the SM-like Higgs state consistent with being a light electroweak
doublet scalar favors that the $SU(2)_L \otimes U(1)_Y$ symmetry is
linearly realized in the effective theory which describes the 
indirect NP effects at LHC
energies~\cite{Buchmuller:1985jz,DeRujula:1991se,
  Hagiwara:1993ck,GonzalezGarcia:1999fq, Passarino:2012cb}.  Except
for total lepton number violating effects, the lowest order operators
which can be built are of dimension six. The coefficients of these
dimension--six operators parametrize our ignorance of the 
NP  and they must be determined using all available data.

An important corollary of this approach is that the modifications of
the couplings of the Higgs field to the electroweak gauge bosons are
related to those of the triple electroweak gauge--boson vertices in a
model independent fashion~\cite{us1,usnew}.  In this letter, we show
that, because of this relation, the analysis of the Higgs boson
production data at LHC and Tevatron is able to furnish bounds on the
related TGCs which are complementary to the direct study of these
couplings in gauge boson production.

More specifically, assuming that the $SU(3)_c \otimes SU(2)_L \otimes
U(1)_Y$ symmetry is realized linearly we can write the lowest order effective
Lagrangian for the departures of the SM as
\begin{equation}
{\cal L}_{\rm eff} =  \sum_n \frac{f_n}{\Lambda^2} {\cal O}_n \;\; ,
\label{l:eff}
\end{equation}
where the dimension--six operators ${\cal O}_n$ involve gauge--bosons,
the Higgs--boson, and/or fermionic fields with couplings $f_n$ and where
$\Lambda$ is a characteristic scale.

{ Restricting to $P$ and $C$--even operators, there are 29
dimension--six operators relevant to the study of the Higgs
couplings~\cite{usnew} barring flavor structure and Hermitian
conjugations. Eight of these modify the Higgs couplings to the
electroweak gauge bosons plus one operator containing
Higgs couplings to gluons and one affecting only the Higgs
self couplings. Three affect only
the Higgs couplings to fermions while the remaining  modify both
the fermionic couplings to the Higgs boson as well as the fermion couplings
to the gauge bosons.  Triple electroweak gauge couplings of on-shell
$W$'s are modified by four of these operators, as well as, by one operator 
that only involves the electroweak gauge--boson self--couplings, 
${\cal O}_{WWW}$ (see Eq.~(\ref{ourope})).

The use of the equations of motion  eliminates 
redundant operators from ${\cal L}_{\rm eff}$.  Moreover,
the coefficients of  several operators are strongly constrained by the
precision electroweak measurements which have helped us to establish
the SM such as $Z$ properties at the pole, $W$ decays, low-energy
$\nu$ scattering, atomic parity violation, flavor changing neutral
currents, parity violation in Moller scattering, and $e^+ e^- \to f
\bar{f}$ at LEP2.} For a detailed discussion on the reduction on the
number of parameters in our effective Lagrangian see
Ref.~\cite{usnew}. At the end of the day, the effective Lagrangian
relevant to the analysis of Higgs couplings and TGCs reads
\begin{eqnarray}
{\cal L}_{eff} \!\!\!
&=& \!\!- \frac{\alpha_s v}{8 \pi} \frac{f_g}{\Lambda^2} 
{\cal O}_{GG}
+ \frac{f_{WW}}{\Lambda^2} {\cal O}_{WW}
+ \frac{f_{BB}}{\Lambda^2} {\cal O}_{BB}\nonumber\\
&+& \frac{f_{\Phi,2}}{\Lambda^2} {\cal O}_{\Phi,2}
+ \frac{f_{\rm bot}}{\Lambda^2}  {\cal O}_{d\Phi,33} 
+ \frac{f_{\tau}}{\Lambda^2} {\cal O}_{e\Phi,33} \label{ourleff}\\
&+& \frac{f_{W}}{\Lambda^2} {\cal O}_{W}
+ \frac{f_{B}}{\Lambda^2} {\cal O}_{B}
+ \frac{f_{WWW}}{\Lambda^2} {\cal O}_{WWW} 
\;\; \nonumber
\end{eqnarray}
with 
\begin{eqnarray}
\!\!\!\!&&{\cal O}_{GG} = \Phi^\dagger \Phi \; G_{\mu\nu}^a G^{a\mu\nu}  \;,
\;\;
{\cal O}_{WW} = \Phi^{\dagger} \hat{W}_{\mu \nu} 
 \hat{W}^{\mu \nu} \Phi  \; , \nonumber \\
&& {\cal O}_{BB} = \Phi^{\dagger} \hat{B}_{\mu \nu} 
  \hat{B}^{\mu \nu} \Phi , \;\; 
  {\cal O}_{\Phi,2} = \frac{1}{2} 
\partial^\mu\left ( \Phi^\dagger \Phi \right)
\partial_\mu\left ( \Phi^\dagger \Phi \right) \; , \nonumber \\
&&{\cal O}_{e\Phi,ij}=(\Phi^\dagger\Phi)(\bar L_{i} \Phi e_{R_j}) 
\; ,  \;\;
{\cal O}_{d\Phi,ij}
=(\Phi^\dagger\Phi)(\bar Q_{i} \Phi d_{Rj})\; , \nonumber  \\
&&{\cal O}_W  = (D_{\mu} \Phi)^{\dagger}  
 \hat{W}^{\mu \nu}  (D_{\nu} \Phi) \; , \;\;
 {\cal O}_B  =  (D_{\mu} \Phi)^{\dagger} 
  \hat{B}^{\mu \nu}  (D_{\nu} \Phi)  \; ,  \nonumber \\
&& {\cal O}_{WWW}=
\hbox{Tr}[\hat{W}_{\mu \nu}\hat{W}^{\nu\rho}\hat{W}_{\rho}^{\mu}]
\;\;. 
\label{ourope}
\end{eqnarray}
$\Phi$ is the Higgs doublet with covariant
derivative $D_\mu\Phi= \left(\partial_\mu+i \frac{1}{2} g' B_\mu +
i g \frac{\sigma_a}{2} W^a_\mu \right)\Phi$ and 
$v=246$ GeV is  its vacuum expectation value. $\hat{B}_{\mu \nu} = i
\frac{g'}{2} B_{\mu \nu}$ and $\hat{W}_{\mu\nu} = i \frac{g}{2}
\sigma^a W^a_{\mu\nu}$ with  $SU(2)_L$ ($U(1)_Y$)
gauge coupling $g$ ($g^\prime$) and  Pauli matrices 
$\sigma^a$.

The first eight operators in Eq.~(\ref{ourleff}) contribute to Higgs
interactions with SM gauge--boson, bottom--quarks and tau pairs; see
Refs.~\cite{us1,usnew} for the explicit form of these interactions.

The last three operators in Eqs.~(\ref{ourleff}) and~(\ref{ourope})
contribute to the TGCs $\gamma W^+ W^-$ and $Z W^+W^-$ that can be
parametrized as~\cite{Hagiwara:1986vm}
\begin{eqnarray}
{\cal L}_{WWV} &=& 
 -i g_{WWV} \Big\{ 
g_1^V \Big( W^+_{\mu\nu} W^{- \, \mu} V^{\nu} 
  - W^+_{\mu} V_{\nu} W^{- \, \mu\nu} \Big)\nonumber\\ 
 &+& \kappa_V W_\mu^+ W_\nu^- V^{\mu\nu}
+ \frac{\lambda_V}{m_W^2} W^+_{\mu\nu} W^{- \, \nu\rho} V_\rho^{\; \mu}
 \Big\}
\;\;,
\label{eq:classical}
\end{eqnarray}
where $g_{WW\gamma} = e=g s$ and $g_{WWZ} = g c$ 
with $s$($c$) being the sine (cosine) of the weak mixing angle.
 In general these
vertices involve six C and P conserving
couplings~\cite{Hagiwara:1986vm}.  Notwithstanding,
the
electromagnetic gauge invariance requires that $g_{1}^{\gamma} = 1$,
while the five remaining couplings are related to the dimension--six
operators ${\cal O}_B$, ${\cal O}_W$ and ${\cal O}_{WWW}$ as 
$\kappa^V=1+\Delta \kappa^V$ and $g^Z_1=1+\Delta g^Z_1$ with
\begin{eqnarray}
&&\Delta \kappa_\gamma = 
 \frac{g^2 v^2}{8\Lambda^2}
\Big(f_W + f_B\Big)\,, \;
  \lambda_\gamma = \lambda_Z = 
\frac{3 g^2 M_W^2}{2 \Lambda^2} f_{WWW}\; ,\nonumber \\  
&&\Delta g_1^Z= \frac{g^2 v^2}{8 c^2\Lambda^2}f_W \, ,\; 
\Delta \kappa_Z =   \frac{g^2 v^2}{8 c^2\Lambda^2}
  \Big(c^2 f_W - s^2 f_B\Big)\, . 
\label{eq:wwv}
\end{eqnarray}

In brief, ${\cal O}_B$ and ${\cal O}_W$ contribute both to Higgs physics
and TGCs which means that some changes of the couplings of the Higgs
field to the vector gauge bosons are related to TGCs due to gauge
invariance in a model independent fashion. In the past the bounds from
TGC searches were used to further constrain the Higgs couplings
to electroweak gauge bosons~\cite{GonzalezGarcia:1999fq}.  Conversely,
with the present precision attained on the determination of the Higgs
couplings, it is possible to reverse the argument and derive the
bounds that Higgs data imply on TGCs.

Equation~(\ref{eq:wwv}) implies that only three of the five TGC couplings
are independent in our framework.  They can be chosen to be
$\Delta\kappa_\gamma$, $\lambda_\gamma$, and $\Delta g_1^Z$, while
$\lambda_Z$ and $\Delta\kappa_Z$ are determined by the relations
\begin{eqnarray}
\lambda_Z=\lambda_\gamma\;\;, & & \Delta\kappa_Z
=-\frac{s^2}{c^2}\Delta\kappa_\gamma +\Delta g_1^Z\;. 
\label{eq:lep}
\end{eqnarray}

Routinely, the collider experiments search for anomalous TGC
parametrized as Eq.~(\ref{eq:classical}) through the analysis of
electroweak gauge--boson production.  In most studies one or at most
two couplings at the time are allowed to deviate from the SM
predictions, while the others are fixed to their SM values. In
particular several searches were performed by the LEP, followed by
Tevatron and recently LHC experiments in the constrained framework
determined by the relations in Eq.~(\ref{eq:lep}) , which are usually
denoted as the ``LEP'' scenario.

\begin{figure}
  \centering
  \vskip -24pt
  \includegraphics[width=0.45\textwidth]{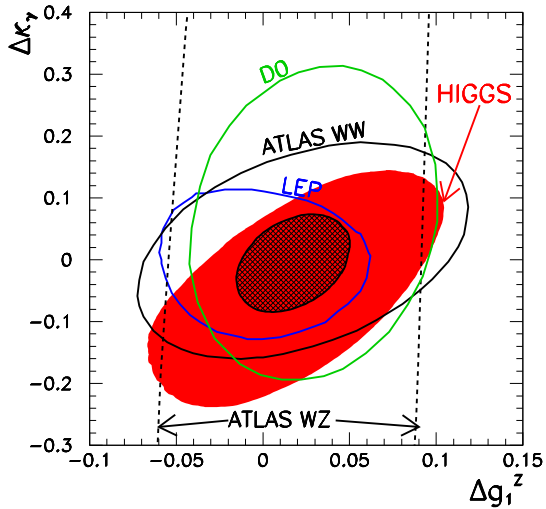}
 \caption{The 95\% C.L. allowed regions (2 d.o.f.) on the plane
   $\Delta\kappa_\gamma \otimes \Delta g^Z_1$ from the analysis of the
   Higgs data from the LHC and Tevatron (filled region) together with the
   relevant bounds from different TGC studies from collider
   experiments as labeled in the figure.  We also show the estimated
   constraints obtainable by combining these bounds (hatched region).}
\label{fig:tgv}
\end{figure}

LEP experiments were sensitive to anomalous TGCs through the $W^+W^-$
and single $\gamma$ and $W$ productions 
which yielded information on both $WWZ$ and $WW\gamma$
vertices~\cite{LEPEWWG}. We depict in Fig.~\ref{fig:tgv} the
bounds obtained in Ref.~\cite{LEPEWWG}
from the combined analysis of the LEP collaborations
in the LEP scenario for $\lambda_\gamma=\lambda_Z=0$.

Tevatron experiments have also set bounds on TGCs from the combination
of $WW$, $WZ$ and $W\gamma$ productions in $p\bar{p}$ collisions. In
the most recent results~\cite{Abazov:2012ze} D\O\ combined these data
sets containing from 0.7 to 8.6 fb$^{-1}$ of integrated luminosity.
CDF has presented results from $WZ$ production~\cite{Aaltonen:2012vu}
with an integrated luminosity of 7.1 fb$^{-1}$ and from $W^+ W^-$ with
3.6 fb$^{-1}$ ~\cite{Aaltonen:2009aa}.  We show in Fig.~\ref{fig:tgv}
the bounds obtained from the D\O\ combined analysis in
Ref.~\cite{Abazov:2012ze} for the LEP scenario.  These bounds were
derived by the experiments for $\lambda_\gamma=\lambda_Z=0$.  Also D\O\
results were obtained assuming a form factor for the anomalous TGC
$\frac{1}{(1 + \frac{\hat{s}}{\Lambda^2})^2}$ with $\Lambda=2$ TeV
\footnote{ It is well known that the introduction of anomalous
  couplings spoils delicate cancellations in scattering amplitudes,
  leading, eventually, to unitarity violation above a certain scale
  $\Lambda$. The way to cure this problem that is used in the
  literature is to introduce an energy dependent form factor that
  dumps the anomalous scattering amplitude growth at high energy.}.

The LHC experiments are providing bounds on
TGCs~\cite{Eboli:2010qd}. ATLAS studied TGCs in
$W^+W^-$~\cite{ATLAS:2012mec}, $WZ$ ~\cite{Aad:2012twa} and $W\gamma$
and $Z\gamma$~\cite{Aad:2013izg} fully leptonic channels at 7 TeV with
an integrated luminosity of 4.6 fb$^{-1}$.  CMS has also constrained
TGCs using 7 TeV data on the leptonic channels in $WW$~\cite{CMS:WWtgc}
with 4.92 fb$^{-1}$, $W\gamma$ and $Z\gamma$ with 5.0
fb$^{-1}$~\cite{CMS:WAtgc}, and $WW$ and $WZ$ productions with two
jets in the final state~\cite{Chatrchyan:2012bd} and 5.0 fb$^{-1}$.
We present in Fig.~\ref{fig:tgv} the most sensitive results from the
LHC searches in the LEP scenario, {\em i.e.}  the $WW$ and $WZ$
studies from ATLAS~\cite{ATLAS:2012mec,Aad:2012twa} (these bounds were
derived by ATLAS for $\lambda_\gamma=\lambda_Z=0$).  Notice that the
limits on the $WWZ$ vertex from the $WZ$ channel~\cite{Aad:2012twa}
were obtained by a two parameter analysis in terms of $\Delta\kappa_Z$
and $\Delta g_1^Z$ and we expressed these bounds in terms of
$\Delta\kappa_\gamma$ and $\Delta g_1^Z$ using Eq.~(\ref{eq:lep}).
Results on $W\gamma$ searches from both ATLAS and
CMS~\cite{Aad:2013izg,CMS:WAtgc} are only sensitive to $WW\gamma$,
{\em i.e.} to $\Delta\kappa_\gamma$ and $\lambda_\gamma$, leading thus
to horizontal bands in Fig.~\ref{fig:tgv}. However they are still
weaker than the bounds shown from $WW$ and $WZ$ productions.  All LHC
bounds in Fig.~\ref{fig:tgv} were obtained without use of form
factors.

We now turn our attention to TGC bounds from Higgs data. In
Ref.~\cite{usnew} an analysis of the latest Higgs data from the LHC
and Tevatron collaborations has been recently updated in this
framework to constrain the relevant dimension-six operators in
Eq.~(\ref{ourleff}):
$f_g,
f_{WW}, f_{BB}, f_{\Phi,2}, f_W, f_B, f_{\rm bot}, \mbox{and } f_\tau$. 
There, it is shown that the inclusion of the fermionic 
operators, $f_{\rm bot} \mbox{ and } f_\tau$,
has a negligible effect on the constraints on $f_W$ and $f_B$.
Thus, for simplicity we show here the results when the fermionic
interactions are set to the SM values, and then the analysis
is made over the six dimensional space spanned by $f_g,
f_{WW}, f_{BB}, f_{\Phi,2}, f_W, \mbox{and } f_B$.
In the analysis,
Eq.~(\ref{eq:wwv}) allows us
to translate the constraints on $f_W$ and $f_B$ from this analysis to
bounds on $\Delta\kappa_\gamma$, $\Delta\kappa_Z$ and $\Delta g_1^Z$
of which only two are independent. We show the results of the fitting
to the Higgs data only in Fig.~\ref{fig:tgv} where we plot the 95\% C.L.
allowed region in the plane $\Delta\kappa_\gamma \otimes \Delta g_1^Z$
{\sl after marginalizing over the other four parameters on the
  Higgs analysis}, $f_g,f_{WW},f_{BB}$ and $f_{\Phi,2}$.  In other
words, we define
\begin{eqnarray}
&&\Delta\chi^2_{H}(\Delta\kappa_\gamma,\Delta g^Z_1) 
=\\
&&\ \ {\rm min}_{f_g,f_{WW},f_{BB},f_{\Phi,2}}
\Delta\chi^2_H(f_g,f_{WW},f_{BB},f_{\Phi,2},f_B,f_W) \;\;.
\nonumber
\end{eqnarray}
So we are not making any additional assumption about the
coefficients of the { six} operators which contribute to the Higgs analysis. 
Notice also that these bounds obtained from the Higgs data are
independent of the value of $\lambda_\gamma=\lambda_Z$. We define the
two-dimensional 95\% C.L. allowed region from the condition
$\Delta\chi^2_{H}(\Delta\kappa_\gamma,\Delta g^Z_1)\leq 5.99$.

Clearly the present Higgs physics bounds on $\Delta\kappa_\gamma
\otimes \Delta g^Z_1$ in Fig.~\ref{fig:tgv} exhibit a non-negligible
correlation.  This stems from the strong correlation imposed on the
high values of $f_W$ and $f_B$ from their tree level contribution to
$Z\gamma$ data, a correlation which is indubitably translated to the
$\Delta\kappa_\gamma \otimes \Delta g^Z_1$ plane.  { The 90\%
C.L. 1 d.o.f. allowed ranges read}
\begin{eqnarray}
-0.047\leq\Delta g^Z_1\leq 0.089\;,&&
-0.19\leq\Delta\kappa_\gamma\leq0.099 \\
{\rm which \; imply}&&\;-0.019\leq\Delta\kappa_Z\leq0.083 \nonumber 
\end{eqnarray}

Figure~\ref{fig:tgv} also shows that the present constraints on
$\Delta\kappa_\gamma \otimes \Delta g^Z_1$ from the analysis of Higgs
data exhibit a correlation which is different from the correlation in the bounds
 from direct TGC studies
  at colliders.  Nevertheless, what is most important is that this
figure illustrates the complementarity of the bounds on NP effects
originating from the analysis of Higgs signals and from studies of the
gauge--boson couplings. To estimate the potential of this
complementarity we combine the present bounds derived from Higgs data
with those from the TGC analysis from LEP, Tevatron and LHC shown in
Fig.~\ref{fig:tgv}.  In order to do so we reconstruct an approximate
Gaussian $\chi^2_i(\Delta\kappa_\gamma,\Delta g^Z_1)$ which reproduces
each of the 95\% C.L. regions for the TGC analysis in the figure
($i=$LEP, D0, ATLAS WW, ATLAS WZ), {\em i.e.}  we obtain the best fit
point and two-dimensional covariance matrix which better reproduce the
curve from the condition $\chi^2_i=5.99$.  So we write
\begin{equation}
   \chi^2_{\rm comb}=\chi^2_H(\Delta\kappa_\gamma,\Delta
   g^Z_1)+{\displaystyle \sum_i} \chi^2_i(\Delta\kappa_\gamma,\Delta
   g^Z_1) \;\;.
\end{equation}
The combined 95\% C.L. region in Fig.~\ref{fig:tgv} is obtained with the condition
$\Delta\chi^2_{\rm comb}\leq 5.99$. 
{ The combined 90\% C.L.\\ 1 d.o.f. allowed ranges read }
\begin{eqnarray}
-0.005\leq\Delta g^Z_1\leq 0.040,&
-0.058\leq\Delta\kappa_\gamma\leq0.047 & \\
{\rm which \; imply}&\;-0.004\leq\Delta\kappa_Z\leq0.040 & .\nonumber 
\end{eqnarray}

Summarizing, the present data on the Higgs-like particle are consistent
with the assumption that the observed state belongs to a light
electroweak doublet scalar and that the $SU(2)_L \otimes U(1)_Y$
symmetry is linearly realized, as demonstrated in
Ref.~\cite{usnew}. Under this assumptions indirect NP effects
associated with the EWSB sector can be written in terms of an effective
Lagrangian whose lowest order operators are of dimension six. The
coefficients of these dimension--six operators parametrize our
ignorance of these effects and our task
at hand is to determine them using all the available data.  In this
general framework the modifications of the couplings of the Higgs
field to electroweak gauge bosons are related to the anomalous triple
gauge--boson vertex.  In this note, we have shown that at present, the
analysis of the Higgs boson production data at the LHC and Tevatron is
able to furnish bounds on the related TGCs
which are complementary to those obtained from direct triple
gauge--boson coupling analysis. In the near future the LHC
collaborations will release their analysis of TGCs with the largest
statistics of the 8 TeV run. The combination of those with the present
results from Higgs data has the potential to furnish the strongest
constraints on NP effects on the EWSB sector.
\vskip 0.5cm 

J.G-F is grateful to the CERN Theory group for their hospitality.
This work is supported 
by Conselho
Nacional de Desenvolvimento Cient\'{\i}fico e Tecnol\'ogico (CNPq) and
by Funda\c{c}\~ao de Amparo \`a Pesquisa do Estado de S\~ao Paulo
(FAPESP),
by USA-NSF grant PHY-09-6739,d 
by CUR Generalitat de Catalunya 
grant 2009SGR502  
by MICINN FPA2010-20807 and consolider-ingenio 2010 program
CSD-2008-0037 and by EU grant FP7 ITN INVISIBLES (Marie Curie Actions
PITN-GA-2011-289442).  
J.G-F is further supported by ME FPU
grant AP2009-2546. 


\begin{thebibliography}{9}

\bibitem{Englert:1964et}
F.~Englert and R.~Brout,
Phys.\ Rev.\ Lett. {\bf 13}, 321 (1964); 
P.~W. Higgs,
Phys.\ Rev.\ Lett. {\bf 13}, 508 (1964); 
P.~W. Higgs,
Phys.\ Lett. {\bf 12}, 132 (1964); 
G.~Guralnik, C.~Hagen, and T.~Kibble,
Phys.\ Rev.\ Lett. {\bf 13}, 585 (1964); 
P.~W. Higgs,
Phys.\ Rev. {\bf 145}, 1156 (1966); 
T.~Kibble,
Phys.\ Rev. {\bf 155}, 1554 (1967).

\bibitem{discovery} 
  S.~Chatrchyan {\it et al.}  [CMS Collaboration],
  Phys.\ Lett.\ B {\bf 716} (2012) 30
  [arXiv:1207.7235 [hep-ex]].
G.~Aad {\it et al.}  [ATLAS Collaboration],
  Phys.\ Lett.\ B {\bf 716} (2012) 1
  [arXiv:1207.7214 [hep-ex]].


\bibitem{us1} 
  T.~Corbett, O.~J.~P.~Eboli, J.~Gonzalez-Fraile and M.~C.~Gonzalez-Garcia,
  Phys.\ Rev.\ D {\bf 86}, 075013 (2012)
  [arXiv:1207.1344 [hep-ph]].

\bibitem{usnew} 
  T.~Corbett, O.~J.~P.~Eboli, J.~Gonzalez-Fraile and
  M.~C.~Gonzalez-Garcia, 
  Phys.\ Rev.\ D {\bf 87}, 015022
  (2013) [arXiv:1211.4580 [hep-ph]]. This analysis has been 
  updated including the results presented by the LHC collaborations
  in March 2013. See also \url{http://hep.if.usp.br/Higgs}. 

\bibitem{couplings} 
For other post-discovery analyses of the $X$ particle couplings, see also: 
J.~Ellis and T.~You,
 JHEP {\bf 1209}, 123 (2012)
  [arXiv:1207.1693 [hep-ph]];
  I.~Low, J.~Lykken and G.~Shaughnessy,
 Phys.\ Rev.\ D {\bf 86}, 093012 (2012)
  [arXiv:1207.1093 [hep-ph]];
P.~P.~Giardino, K.~Kannike, M.~Raidal and A.~Strumia, 
 Phys.\ Lett.\ B {\bf 718}, 469 (2012)
  [arXiv:1207.1347 [hep-ph]];
M.~Montull and F.~Riva,
  JHEP {\bf 1211}, 018 (2012)
  [arXiv:1207.1716 [hep-ph]];
J.~R.~Espinosa, C.~Grojean, M.~Muhlleitner and M.~Trott,
 JHEP {\bf 1212}, 045 (2012)
  [arXiv:1207.1717 [hep-ph]];
D.~Carmi, A.~Falkowski, E.~Kuflik, T.~Volansky and J.~Zupan, 
 JHEP {\bf 1210}, 196 (2012)
  [arXiv:1207.1718 [hep-ph]];
S.~Banerjee, S.~Mukhopadhyay and B.~Mukhopadhyaya,
JHEP {bf 10} (2012) 062,
[arXiv:1207.3588 [hep-ph]];
F.~Bonnet, T.~Ota, M.~Rauch and W.~Winter,
 Phys.\ Rev.\ D {\bf 86}, 093014 (2012)
  [arXiv:1207.4599 [hep-ph]];
T.~Plehn and M.~Rauch,
 Europhys.\ Lett.\  {\bf 100}, 11002 (2012)
  [arXiv:1207.6108 [hep-ph]];
A.~Djouadi,
arXiv:1208.3436 [hep-ph];
B.~Batell, S.~Gori and L.~T.~Wang,
JHEP {\bf 1301}, 139 (2013)
  [arXiv:1209.6382 [hep-ph]];
G.~Moreau,
  Phys.\ Rev.\ D {\bf 87}, 015027 (2013)
  [arXiv:1210.3977 [hep-ph]];
G.~Cacciapaglia, A.~Deandrea, G.~D.~La Rochelle and J-B.~Flament,
arXiv:1210.8120 [hep-ph];
A.~Azatov and J.~Galloway,
  Int.\ J.\ Mod.\ Phys.\ A {\bf 28}, 1330004 (2013)
  [arXiv:1212.1380 [hep-ph]].






\bibitem{Giardino:2013bma} 
For results using the data released in Moriond 2013 see:
  A.~Falkowski, F.~Riva and A.~Urbano,
  arXiv:1303.1812 [hep-ph];
  J.~Ellis and T.~You,
  arXiv:1303.3879 [hep-ph];
   P.~P.~Giardino, K.~Kannike, I.~Masina, M.~Raidal, and A.~Strumia,
  arXiv:1303.3570 [hep-ph];
   T.~Alanne, S.~Di Chiara and K.~Tuominen,
  arXiv:1303.3615 [hep-ph].
  A.~Djouadi and  G.~Moreau,
  arXiv:1303.6591 [hep-ph].

\bibitem{effective} S.\ Weinberg, {\sl Physica} {\bf 96A}, 327 (1979).


\bibitem{Buchmuller:1985jz}
W.~Buchmuller and D.~Wyler,
 Nucl.\ Phys.\ {\bf B268}, 621 (1986); 
C.~N. Leung, S.~Love, and S.~Rao,
\newblock Z.\ Phys.\ {\bf C31}, 433 (1986); 
  B.~Grzadkowski, M.~Iskrzynski, M.~Misiak and J.~Rosiek,
  JHEP {\bf 1010}, 085 (2010)
  [arXiv:1008.4884 [hep-ph]].

\bibitem{DeRujula:1991se}
A.~De~Rujula, M.~Gavela, P.~Hernandez, and E.~Masso,
 Nucl.\ Phys.\  {\bf B384}, 3 (1992).

\bibitem{Hagiwara:1993ck}
K.~Hagiwara, S.~Ishihara, R.~Szalapski, and D.~Zeppenfeld,
 Phys.\ Rev.\ {\bf D48}, 2182 (1993);
%
K.~Hagiwara, R.~Szalapski, and D.~Zeppenfeld,
Phys.\ Lett.\ {\bf B318}, 155 (1993) [arXiv:hep-ph/9308347];
K.~Hagiwara, S.~Matsumoto, and R.~Szalapski,
Phys.\ Lett.\ {\bf B357}, 411 (1995) [arXiv:hep-ph/9505322].


\bibitem{GonzalezGarcia:1999fq}
M.~Gonzalez-Garcia,
 Int.\ J.\ Mod.\ Phys.\ {\bf A14}, 3121 (1999), arXiv:hep-ph/9902321;
  F.~de Campos, M.~C.~Gonzalez-Garcia and S.~F.~Novaes,
  Phys.\ Rev.\ Lett.\  {\bf 79}, 5210 (1997)
  [hep-ph/9707511];
  M.~C.~Gonzalez-Garcia, S.~M.~Lietti and S.~F.~Novaes,
  Phys.\ Rev.\ D {\bf 57}, 7045 (1998)
  [hep-ph/9711446];
  O.~J.~P.~Eboli, M.~C.~Gonzalez-Garcia, S.~M.~Lietti and S.~F.~Novaes,
  Phys.\ Lett.\ B {\bf 434}, 340 (1998)
  [hep-ph/9802408];
  M.~C.~Gonzalez-Garcia, S.~M.~Lietti and S.~F.~Novaes,
  Phys.\ Rev.\ D {\bf 59}, 075008 (1999)
  [hep-ph/9811373];
  O.~J.~P.~Eboli, M.~C.~Gonzalez-Garcia, S.~M .Lietti and S.~F.~Novaes,
  Phys.\ Lett.\ B {\bf 478}, 199 (2000)
  [hep-ph/0001030].


\bibitem{Passarino:2012cb} 
G.~Passarino,
  Nucl.\ Phys.\ B {\bf 868}, 416 (2013)
  [arXiv:1209.5538 [hep-ph]].

\bibitem{Hagiwara:1986vm}
K.~Hagiwara, R.~Peccei, D.~Zeppenfeld, and K.~Hikasa,
Nucl.\ Phys.\ {\bf B282}, 253 (1987).

\bibitem{LEPEWWG} The LEP collaborations and The LEP TGC Working group
LEPEWWG/TGC/2003-01

\bibitem{Abazov:2012ze} 
  V.~M.~Abazov {\it et al.}  [D0 Collaboration],
  Phys.\ Lett.\ B {\bf 718}, 451 (2012)
  [arXiv:1208.5458 [hep-ex]].

\bibitem{Aaltonen:2012vu} 
  T.~Aaltonen {\it et al.}  [CDF Collaboration],
  Phys.\ Rev.\ D {\bf 86}, 031104 (2012)
  [arXiv:1202.6629 [hep-ex]].
  
\bibitem{Aaltonen:2009aa} 
  T.~Aaltonen {\it et al.}  [CDF Collaboration],
  Phys.\ Rev.\ Lett.\  {\bf 104}, 201801 (2010)
  [arXiv:0912.4500 [hep-ex]].

\bibitem{Eboli:2010qd}
For a study of the sensitivy to TGC in the early LHC runs see  
  O.~J.~P.~Eboli, J.~Gonzalez-Fraile and M.~C.~Gonzalez-Garcia,
  Phys.\ Lett.\ B {\bf 692}, 20 (2010)
  [arXiv:1006.3562 [hep-ph]] and references therein.
  
\bibitem{ATLAS:2012mec} 
  G.~Aad {\it et al.}  [ATLAS Collaboration],
  arXiv:1210.2979 [hep-ex].
  
\bibitem{Aad:2012twa} 
  G.~Aad {\it et al.}  [ATLAS Collaboration],
  Eur.\ Phys.\ J.\ C {\bf 72}, 2173 (2012)
  [arXiv:1208.1390 [hep-ex]].
  
\bibitem{Aad:2013izg} 
  G.~Aad {\it et al.}  [ The ATLAS Collaboration],
  arXiv:1302.1283 [hep-ex].
  
\bibitem{CMS:WWtgc}
  CMS Collaboration,
  \url{
   https://twiki.cern.ch/twiki/bin/view/CMSPublic/PhysicsResultsSMP12005
  }

\bibitem{CMS:WAtgc}
  CMS Collaboration,
  \url{
  https://twiki.cern.ch/twiki/bin/view/CMSPublic/PhysicsResultsEWK11009
  }
  
\bibitem{Chatrchyan:2012bd} 
  S.~Chatrchyan {\it et al.}  [CMS Collaboration],
  Eur.\  Phys.\  J.\  C {\bf 73}, 2283 (2013)
  [arXiv:1210.7544 [hep-ex]].

  
\end{thebibliography}
\end{document}